\documentclass[showpacs,preprintnumbers,amsmath,amssymb,prl,twocolumn,
superscriptaddress]{revtex4}

\usepackage{graphicx}
\usepackage{dcolumn}
\usepackage{bm}
\def\comment#1{}

\def\comment#1{}

\newcommand{\nc}{\newcommand}
\nc{\beq}{\begin{eqnarray}}
\nc{\eeq}{\end{eqnarray}}
\nc{\scs}{\scriptstyle}
\nc{\setval}{\fmfset{wiggly_len}{3mm} \fmfset{arrow_len}{1.5mm}
    \fmfset{arrow_ang}{13} \fmfset{dash_len}{1.5mm}\fmfpen{0.125mm}
    \fmfset{dot_size}{2thick}}

\begin{document}

\title{
Defect Melting Models for Cubic Lattices and\\
Universal Laws for Melting Temperatures}

\author{Hagen Kleinert}
\email{kleinert@physik.fu-berlin.de}
\affiliation{Institut f\"ur Theoretische Physik,
Freie Universit\"at Berlin, Arnimallee 14, D-14195 Berlin, Germany}
\author{Ying Jiang}
\email{jiang@physik.fu-berlin.de}
\affiliation{Institut f\"ur Theoretische Physik,
Freie Universit\"at Berlin, Arnimallee 14, D-14195 Berlin, Germany}

\date{Received \today}

\begin{abstract}
We set up simple harmonic lattice models
for  elastic fluctuations in bcc and fcc lattices
and the excitation of dislocations
and disclinations. From these
we derive, in  a lowest approximation,
universal formulas  which
 predict
 melting temperatures
in good agreement with the
experiments. This new theory is more precise
than Lindemann's rule by factor 2, and more predictive,
since the size of the Lindemann number has to be fixed by
experiments.
In addition, our theory  allows for systematic
improvements.

\end{abstract}

\maketitle

Melting transitions \cite{shoc1,kleinert1,kleinert2,kleinert118} are
important phenomena
of both technical and theoretical interest \cite{zhang1,%
dash1,trumper,kleinert4,nelson,wang,kleinert179,kleinert183,kleinert189}.
A hundred years ago, Sutherland \cite{sutherland1} found an
empirical rule, that the product $bT_mM^{\frac16}$ is nearly constant
for metals, where $T_m$ is the melting temperature, $b$  the mean
coefficient of expansion, and $M$ the atomic mass. In the following
year, he advanced a kinetic theory of solids  \cite{sutherland2}, in which
he postulated that melting would occur when the space between  atoms
reaches a certain value relative to the atomic diameter, so that
atoms are able to escape from the imprisonment by their neighbors.
He carried out a hard-sphere simulation experiment by shaking a box
containing one layer of marbles, and noting by removing marbles one
at a time that the apparent solid-liquid transition occurs when a free
volume of 25-33\% is reached.
From
the vibration amplitude and the thermal kinetic energy,
together with the
above  empirical melting rule, Sutherland \cite{sutherland2} calculated
the period of vibrations of metal atoms at the melting temperature, and found
that the ratio of vibration amplitude to atomic spacing is nearly the same
for all elements at melting.

In 1910, Lindemann \cite{lindemann1} combined Sutherland's
ideas
with the Debye theory
of specific heat
in solids, and
derived his famous
rule according to which a
quasi-universal parameter, the Lindemann number,
should  be the same for all melting transitions.
The number is usually
stated  in the form
\[
L = \theta v^{1/3}(M/T_m)^{1/2}
\]
where $ \theta$ is the Debye temperature  and $ v$ the volume per atom.
Subsequent extensive tests of this rule were carried out by
Gschneidner \cite{gschneidner1}, Ross \cite{ross1}, and Crawford \cite{crawford1}
who found,
however, that Lindemann's rule is not very reliable.
First, the Debye
temperature cannot be precisely specified
since the Debye theory of specific heat is itself approximate.
Second, extracting $\theta$ from the data by a best fit of the Debye theory,
 the Lindemann number varies considerably from material to material, as
emphasized  by Wallace \cite{wallace1}.

The most unsatisfactory feature
of the Lindemann rule is that
its size is unknown and has to be
extracted from an average
of many melting transitions.
 Thus,
the Sutherland-Lindemann approach to melting has remained
a purely phenomenological
rough description
of the transition.
 In spite of its roughness,
it is often used even today in many works to estimate melting temperature,
this being due to
a lack of a more precise melting theory.

During the past two decades,
several
 other melting theories have been developed.
 One is based
on density functionals \cite{curtin1,denton1},
and has provided us with an alternative
insight
into
the
melting transition.
This work,  however,
describes satisfactorily
only the long-wave length limit of fluctuations, and
 neglects the important role of
the lattice structure, thus requiring essential corrections,
which sometimes have been inserted these heuristically
 \cite{vaulina,burakovsky,cho}.
Another theory of melting is based on a Landau expansion in a symmetric
tensor order parameter of rank four \cite{hess}. This theory
is purely phenomenological and thus only
descriptive, not predictive.

For a  proper description for the
melting transition, it seems necessary to
 incorporate information of
the lattice structure and the associated crystal defects
into the theory. About 15 years ago,
one of the authors (HK) did this successfully by constructing  lattice
models of the melting transition
\cite{kleinert2,kleinert118}.
The important progress of these models  \cite{kleinert93}
was that they started out from the lattice version of the elastic
lattice energy which allow
for the thermal creation and annihilation of
dislocations and disclinations
and keeps track of
their
lattice elastic energy
 \cite{kleinert93}.
They are included  by means of discrete-valued defect gauge fields,
the disclinations being
 essential for explaining the first-order
nature of transition \cite{kleinert120},
which otherwise would
be of second order, as
in the vortex-line induced
$ \lambda $-transition of  superfluid helium.

In this respect, the older models
went beyond those of
more recent
authors \cite{burakovsky,chou} who considered only
the proliferation of dislocation lines
which, if
 properly treated, would have
 produced only second-order phase transitions
\cite{kleinert87},
thus being unable
to describe the melting transition.

A simple universal melting formula
was obtained from the above lattice defect models
in a lowest-order approximation
which combines the properties of the high-temperature
expansion of the defect contributions
with the low-temperature expansion of the free energy density. From the
intersection of these two curves, one can obtain
a formula for the melting temperature
in accordance with Lindemann's rule,
but with a {\em prediction\/}
of the absolute
size of the Lindemann number \cite{kleinert2}.

This successful theory
has, however, one important drawback:
For simplicity, it was constructed only for the
physically somewhat exotic
case of simple cubic lattices.
This makes it strictly applicable only
to very few materials---there are almost no simple cubic lattices in nature.
If the derived formulas
are applied to other cubic crystal structures
such as body center cubic (bcc) lattices and face center cubic (fcc) lattices,
the accuracy must  necessarily be bad.
Hence there is an obvious need to generalize the theory
to physically more prolific crystal structures.

The purpose of this letter is to present such a
generalization
 of the melting model to fcc and bcc lattices.
When generalizing the previous melting model to other cubic lattices,
 We shall see
in the high-temperature limit,
that for all cubic crystals the free energy density is
only related to the elastic constants of the crystals and does not depend on
the lattice structure of the system.
 This is in contrast to the low-temperature limit,
where the
free energy density involves integrals over the lattice momenta $K_{i}$ and $%
\overline{K}_{i}$ which strongly depend on the detailed lattice structure of the
crystals.
It is therefore important to give a proper description of the low-temperature
limit of the free energy density
for different lattice structures.

The lattice model
for the cubic crystals
will be based on the elastic energy
\begin{eqnarray}
E &=&\frac{a^{3}}{n}A\sum_{x} \left\{ \frac{1}{4}\!\sum_{l\neq m}\left[
\nabla _{(l)}u_{(m)}+\nabla _{(m)}u_{(l)}\right] ^{2}\right.\nonumber \\&+&\left.C\sum_{l}\left[ \nabla
_{(l)}u_{(l)}\right] ^{2}\!\!+\!\frac{B}{2A}\left[ \sum_{l}\nabla _{(l)}u_{(l)}\left(%
\vec{x}\!-\!\vec{e}_{(l)}\frac{a}{2}\right)\!\right] ^{2}\!\right\}  \nonumber \\
&=&-\frac{a^{3}}{n}A\sum_{x,i,j,l,n}\frac{1}{2}u_{i}(x)\Bigg[
e_{(l)}^{i}e_{(l)}^{j}\sum_{m}\overline{\nabla}_{(m)}\nabla
_{(m)}
\nonumber \\&&~~~~~~~~~~~~~~+
2(C-1)e_{(l)}^{i}e _{(l)}^{j}\nabla _{(l)}\overline{\nabla}_{(l)}
\nonumber \\&&~~~~~~~~~~~~~~+\left(\frac{B}{A}+1\right)e_{(l)}^{i}\nabla _{(l)}
\overline{\nabla} _{(n)}e_{(n)}^{j}\Bigg] u_{j}(x),
\label{eefcc1}
\end{eqnarray}%
where $n$ is the  number of atoms  per unit cell,
$\vec{e}_{(l)}$ are {\it oriented link vectors} of cubic
lattice pointing from any lattice site
to its nearest neighbors along the {\it positive} direction,
 and $u_i$
denote the Cartesian displacement field,
$u_{(l)}=u_i e^i _{(l)}$ their components
along the link directions.
For fcc lattices, there are six link vectors:
\begin{eqnarray}
&&\!\!\!\!\!\!\!\!\!\!\!\!\!\!\!\!\vec{e}_{(1)}=(1,1,0),~\,~\;\vec{e}_{(2)}=(1,0,1),~\,\;\;\;\vec{e}_{(3)}=(0,1,1)
\nonumber \\
&&\!\!\!\!\!\!\!\!\!\!\!\!\!\!\!\!\vec{e}_{(4)}=(1,-1,0),\;\vec{e}_{(5)}=(0,1,-1),\;\;\vec{e}_{(6)}=(-1,0,1),
\end{eqnarray}%
each lattice site ${\bf x}$ having twelve
nearest neighbors at ${\bf x}\pm \vec{e}_{(l)}{a}/{2}$. For bcc
lattices,
there are four link vectors
\begin{eqnarray}
\vec{e}_{(1)}=(1,1,1),\;~~\,\vec{e}_{(2)}=(1,1,-1),
\nonumber \\
\vec{e}_{(3)}=(1,-1,1),\;
\vec{e}_{(4)}=(-1,1,1),
\label{1}\end{eqnarray}
and each lattice site possesses eight nearest neighbors.
The symbols
$\nabla _{(l)}$ and
$\overline\nabla _{(l)}$
denote lattice derivative, defined by
$\nabla _{(l)}f({\vec x})=
\left[f\left({\vec x}+\vec{e}_{(l)}{a}/2\right)-f(%
{\vec x})\right]/b$ and $
\overline\nabla _{(l)}f({\vec x})=
\left[
f({\vec x})
-f\left({\vec x}-\vec{e}_{(l)}{a}/{2}\right)
\right]/b  ,
$
where $b$ is the distance between the nearest neighbors. After some algebra
calculation,
 the
prefactors $A$, $B$, and $C$
are related to
the usual elastic constant of the crystal
 $\mu $, $\lambda $ and $\xi $
defined by the continuum elastic energy
\begin{eqnarray}
E=\mu \int d^3x\left[ \sum _{i\neq j}u
_{ij} ^2 + \xi \sum _i u _{ii} ^2 +\frac{\lambda}{2 \mu}\left(\sum _i u _{ii}\right)^2\right],
\label{2}\end{eqnarray}
and can be expressed for  fcc as follows:
\begin{eqnarray}
A\!=\!\frac{\mu }{8}(2\xi -1),\;
B\!=\!\frac{\lambda\! +\!2\mu (\xi -1)}{8},\;
C\!=\!1\!-\!8%
\frac{\xi -1}{2\xi -1},
\label{@elen}\end{eqnarray}%
and for bcc as
\begin{eqnarray}
A=\frac{3}{16}\mu \xi ,\;\;\;B=\frac{3}{16}(\lambda +\mu \xi -\mu ),\;\;\;C=%
\frac{2-\xi }{\xi },
\end{eqnarray}

For very low temperatures, the atomic positions deviate very little from
those of an ideal crystal. It is therefore suggestive to use these small
deviations for defining the displacement field. But this definition can be
consistent only for a small time span. Due to fluctuations, thermals as well
as quantum, the atoms are capable of exchanging positions with their neighbors
and migrate, after a sufficiently long time, through the entire crystal. This
process of self-diffusion makes it impossible to specify the displacement
field uniquely. Thus, as a matter of principle, the displacement field is
undetermined up to an arbitrary lattice vector, it is impossible to say
whether an atom is displaced by $u_{(l)}(x)$ or by
\[
u_{(l)}({\vec x})+ b N_{(l)}({\vec x})
\]
where $N_{(l)}({\vec x})$ is the jumping field with integer value. Because of
this, we should introduce an extra sum over an integer-valued field
$n _{(lm)}(x)$ with a gauge
fixing term $\Phi[n _{(lm)}]$ in the corresponding partition function,
thus the partition function of the defect melting model reads
\begin{eqnarray}
Z\!&=&\!\sum _{n _{(lm)}}\prod _{x,i}\left[\int \frac{du_i(x)}{a}
\right] \Phi[n _{(lm)}] \nonumber \\
\!& &\! \times \exp \Bigg \{\!-\frac{a^3 A}{nk_BT}\sum_x \Bigg[ \frac12\sum _{l<m}
\Big(\nabla_{(l)}u_{(m)}+\nabla_{(m)}u_{(l)} \nonumber \\
\!& &\!- b(n_{(lm)}+n_{(ml)})\Big)^2 + C\sum _{l}\left(\nabla _{(l)}u _{(l)}
- bn_{(ll)}\right)^2 \nonumber \\
\!& &\!\! + \frac{B}{2A}\Big(\!\sum _{l}\!\big[\!\nabla _{(l)}u_{(l)}(\vec x\! -\vec e_{(l)}\!\frac a2\!) \nonumber \\
\! & & - b n_{(ll)}(\vec x\! -\vec e_{(l)}\!\frac a2) \big] \Big)^2 \Bigg]  \Bigg \}
\label{latticepartition}
\end{eqnarray}
The partition function is invariant under the following defect gauge
transformations
\[
u _{(l)}  \rightarrow  u _{(l)}(x) + b N_{(l)}(x) ,~
n _{(lm)} \rightarrow  n _{(lm)} + \nabla _{(l)} N _{(m)}(x).
\]
This partition function can be expressed in terms
of the stress field $\sigma _{ij}$ by
\begin{eqnarray}
Z\!\! &=&\!\! \left[8\xi^3\left(1\!+\!\frac{3\lambda}{2\mu\xi}\right)\! \right]^{-N/2}\! 
\left(\frac{1}{2\pi\beta}\right)^{3N}\!\prod _{x, i\leq j}\!\left[ \int\! d \sigma_{ij}\right] \nonumber \\
&\times&\!\sum _{n_{(lm)}}\!\Phi[n_{(lm)}]\! \prod _{x,i}\left[\int \frac{du_i(x)}{a}
\right] \exp\Bigg\{\frac{-1}{2\beta}\sum _x\Bigg[\sum_{i<j}\sigma _{ij}^2 \nonumber \\
&+&\!
\frac{1}{2\xi}\sum _i\sigma _{ii}^2-\frac{\lambda}{4\mu \xi^2 +6 \lambda \xi}\Big(\sum _i \sigma _{ii}
\Big)^2\Bigg]\Bigg\} \nonumber \\
&\times&\! \exp\!\Bigg[\!i\pi\!\sum _{x,l,m}\!\sigma _{\!(lm)}\!\Big(\!\nabla _{\!(l)}
\!u _{\!(m)}\! +\! \nabla _{\!(m)}\! u _{\!(l)}\! -2bn _{\!(lm)}\!\Big)\!\Bigg]
,
\label{partition2}
\end{eqnarray}
\vspace{-1mm}

\noindent
where $\sigma _{(lm)}=(\sqrt{3}/16) e_{(l)}^i e_{(m)}^j \sigma _{ij}$ for
bcc lattice and $\sigma _{(lm)}=(\sqrt{2}/16) e_{(l)}^i
e_{(m)}^j \sigma _{ij}$ for fcc lattice, and $\beta=(a^3\mu)/(nk_BT(2\pi)^2)$. In the calculation process from
(\ref{latticepartition}) to (\ref{partition2}), the corresponding expressions
of $A$, $B$, $C$ and $e_{(l)}^i$ are used.
In spite of the
tedious intermediate calculation, the prefactors in the partition functions
(\ref{partition2}) are the same for all cubic lattices.

For the-lowest order approximation, we need only to investigate the
 system in  low-temperature  and high-temperature limits.
It has been shown in the textbook \cite{kleinert2} (see Fig. 12.1 on p. 1084)
that the intersection of the free energies of the
two limiting curves yields good estimates for the melting temperatures
in simple-cubic lattices.

For $T\rightarrow0$, the defect configuration is completely frozen out, and
the partition function has a classical limit.
From the lattice energy (\ref{eefcc1})
 we obtain the classical
partition function of cubic lattice%
\[
Z_{\rm cl} =\prod_{x,i}\left[\int \frac{du_{i}(x)}{a}\right]e^{-E/k_{B}T}
=\sqrt{\frac{2\pi nk_{B}T}{Aa^{3}}}^{3N} \!\!\!
e^{ -N\frac{1}{2}\ell
},
\]
where  the dimensionless parameter $\ell$ is defined by the trace log
\begin{eqnarray}
\ell =\int\limits_{-2\pi/a }^{2\pi/a }\frac{d^{3}ka^3}{(4\pi )^{3}}{\rm tr}\log
{\bf {M}},  \label{ell}
\end{eqnarray}
and
 ${\bf M}$ denotes the matrix
\begin{eqnarray}
M_{ij}\!\!\!&=&\!\!\!4\delta _{ij}a^{2}\!\sum_{m}\!\overline{K}_{(m)}K_{(m)}
\!\!+\!2(C\!-\!1)a^{2}\!%
\sum_{l}\!e_{(l)}^{i}K_{(l)}\overline{K}_{(l)}e_{(l)}^{j}
\nonumber \\&&
\!+\!\left(\!\frac{B}{A}
\!+\!1
\!\right)\!a^{2}\sum_{l,n}\!e_{(l)}^{i}K_{(l)}\overline{%
K}_{(n)}e_{(n)}^{j} \label{m}
\end{eqnarray}
and $K _{(l)}$ and $\overline{K} _{(l)}$ are lattice momenta, for fcc
\begin{eqnarray}
K_{(l)}=\frac{\sqrt{2}}{ai}(e^{i\vec{k}\cdot \vec{e}_{(l)}\frac{a}{2}%
}-1),\;\overline{K}_{(l)}=\frac{\sqrt{2}}{ai}(1-e^{-i\vec{k}\cdot \vec{e}_{(l)}%
\frac{a}{2}}),
\end{eqnarray}%
and for bcc
\begin{eqnarray}
\!K_{(l)}\!=\frac{2i}{\sqrt{3}a}(1\!-\!e^{i\vec{k}\cdot \vec{e}_{(l)}\frac{a%
}{2}}),\overline{K}_{(l)}\!=\frac{2i}{\sqrt{3}a}(e^{-i\vec{k}%
\cdot \vec{e}_{(l)}\frac{a}{2}}\!-\!1).
\end{eqnarray}
Thus we find
the free energy density of the fcc and bcc lattices in
the low-temperature limit%
\begin{eqnarray}
-\frac{f^{T\rightarrow 0}}{k_BT}=\frac{3}{2}\log \left(\frac{2\pi nk_BT}{\mu a^3}
\right)+\frac{3}{2}\log
\frac{\mu }{A}-\frac{1}{2}\ell .  \label{freelowtfcc}
\end{eqnarray}%

In the opposite limit of high temperature,
 the defects are prolific
and the sum over $n _{(lm)}$ can be approximated by integral
 enforcing $\sigma _{ij}\equiv0$. Then the partition function
 (\ref{partition2}) yields
the free energy density
\begin{eqnarray}\!\!\!
-\frac{ f^{T\rightarrow \infty}}{k_B T}\!=\!3\!\log\!
\left(\!\frac{2\pi nk_BT}{\mu a^3}\!\right)
\!-\frac{1}{2}\!\log\!\left[\!8\xi
^3\!\left(1\!+\!\frac{3\lambda}{2\mu \xi}\right)\!\right]\!.
\label{htapp}\end{eqnarray}
From the intersection of this with
the low-temperature approximation
(\ref{freelowtfcc}),
we obtain the melting relation for the fcc and bcc lattices%
\begin{eqnarray}
\frac{a^{3}\mu \xi }{2\pi nk_{B}T_{\rm melt}}=\frac{A}{\mu }\frac{1}{8^{1/3}}%
\displaystyle  \left(
1+\frac{3\lambda }{2\xi \mu }\right)^{-1/3}e^{\ell /3}\;\;\; .
\label{meltfcc}
\end{eqnarray}
The formula is structurally very similar to the Lindemann
rule $L=$ const., but {\em predicts\/}, in addition, the size of $L$ to be
averagely of the order of 126.
Moreover, from our expression, one can recognize clearly that the melting
temperature will be zero when the anisotropic parameter $\xi$ equals to
zero, this is in contrast to recent phenomenological theories of melting
by other authors
\cite{burakovsky}.
The novelty of this formula with respect to
a similar formula
 in Ref.~\cite{kleinert2} [see Eq. (12.13) on p. 1079]
lies in a determination of
trace log term $\ell$ for bcc and fcc lattices.
The resulting melting temperatures from Eq. (\ref{meltfcc}) are
shown in Table 1, where they are compared with experimental numbers and with
the results found from  Lindemann's rule
where  the Lindemann number is initially undetermined and must
 be
extracted from the average over all materials.
Our lowest-order theory has an average
 precision of about
12\%, which is better by a factor 2 than
the 22\%
of the numbers from derived Lindemann's rule.

\begin{center}
{\footnotesize\begin{tabular}{||l|c|c|c|c|c|c|c||}
\hline
Element & $\mu$ & $\lambda$ & $\xi$ & $T_{m,{\rm theor.}}$
& $T_{m,{\rm exp.}}$ & $\kappa (\%) $ & $\delta (\%) $ \\ \hline
Ag (fcc) & \multicolumn{1}{r|}{51.10} & \multicolumn{1}{r|}{97.30} &
\multicolumn{1}{r|}{0.335} & \multicolumn{1}{r|}{1246.5} &
\multicolumn{1}{r|}{1234.0} & \multicolumn{1}{r|}{1} &
\multicolumn{1}{r||}{24} \\
Au (fcc) & \multicolumn{1}{r|}{45.40} & \multicolumn{1}{r|}{169.70} &
\multicolumn{1}{r|}{0.351} & \multicolumn{1}{r|}{1201.4} &
\multicolumn{1}{r|}{1336.2} & \multicolumn{1}{r|}{-10} &
\multicolumn{1}{r||}{6} \\
Ba (bcc) & \multicolumn{1}{r|}{9.50} & \multicolumn{1}{r|}{8.00} &
\multicolumn{1}{r|}{0.242} & \multicolumn{1}{r|}{875.4} &
\multicolumn{1}{r|}{998.0} & \multicolumn{1}{r|}{-12} &
\multicolumn{1}{r||}{22} \\
Ca (fcc) & \multicolumn{1}{r|}{16.30} & \multicolumn{1}{r|}{18.20} &
\multicolumn{1}{r|}{0.294} & \multicolumn{1}{r|}{889.0} &
\multicolumn{1}{r|}{1112.0} & \multicolumn{1}{r|}{-20} &
\multicolumn{1}{r||}{-14} \\
Co (fcc) & \multicolumn{1}{r|}{128.00} & \multicolumn{1}{r|}{160.00} &
\multicolumn{1}{r|}{0.320} & \multicolumn{1}{r|}{1892.7} &
\multicolumn{1}{r|}{1765.0} & \multicolumn{1}{r|}{7} &
\multicolumn{1}{r||}{-2} \\
Cu (fcc) & \multicolumn{1}{r|}{81.80} & \multicolumn{1}{r|}{124.90} &
\multicolumn{1}{r|}{0.314} & \multicolumn{1}{r|}{1299.8} &
\multicolumn{1}{r|}{1356.0} & \multicolumn{1}{r|}{-4} &
\multicolumn{1}{r||}{16} \\
Li (bcc) & \multicolumn{1}{r|}{10.80} & \multicolumn{1}{r|}{12.50} &
\multicolumn{1}{r|}{0.110} & \multicolumn{1}{r|}{450.2} &
\multicolumn{1}{r|}{454.0} & \multicolumn{1}{r|}{-1} &
\multicolumn{1}{r||}{-14} \\
Nb (bcc) & \multicolumn{1}{r|}{28.70} & \multicolumn{1}{r|}{134.00} &
\multicolumn{1}{r|}{1.951} & \multicolumn{1}{r|}{2778.2} &
\multicolumn{1}{r|}{2741.0} & \multicolumn{1}{r|}{1} &
\multicolumn{1}{r||}{-39} \\
Ni (fcc) & \multicolumn{1}{r|}{124.70} & \multicolumn{1}{r|}{147.30} &
\multicolumn{1}{r|}{0.398} & \multicolumn{1}{r|}{2034.9} &
\multicolumn{1}{r|}{1726.0} & \multicolumn{1}{r|}{17} &
\multicolumn{1}{r||}{16} \\
Pb (fcc) & \multicolumn{1}{r|}{14.40} & \multicolumn{1}{r|}{39.20} &
\multicolumn{1}{r|}{0.222} & \multicolumn{1}{r|}{507.4} &
\multicolumn{1}{r|}{600.6} & \multicolumn{1}{r|}{-15} &
\multicolumn{1}{r||}{26} \\
Pd (fcc) & \multicolumn{1}{r|}{71.20} & \multicolumn{1}{r|}{176.10} &
\multicolumn{1}{r|}{0.407} & \multicolumn{1}{r|}{1723.0} &
\multicolumn{1}{r|}{1825.0} & \multicolumn{1}{r|}{-6} &
\multicolumn{1}{r||}{7} \\
Pt (fcc) & \multicolumn{1}{r|}{76.50} & \multicolumn{1}{r|}{250.70} &
\multicolumn{1}{r|}{0.627} & \multicolumn{1}{r|}{2527.3} &
\multicolumn{1}{r|}{2042.0} & \multicolumn{1}{r|}{23} &
\multicolumn{1}{r||}{29} \\
Sr (fcc) & \multicolumn{1}{r|}{9.90} & \multicolumn{1}{r|}{10.30} &
\multicolumn{1}{r|}{0.253} & \multicolumn{1}{r|}{868.0} &
\multicolumn{1}{r|}{1045.0} & \multicolumn{1}{r|}{-17} &
\multicolumn{1}{r||}{11} \\
Ta (bcc) & \multicolumn{1}{r|}{81.80} & \multicolumn{1}{r|}{157.40} &
\multicolumn{1}{r|}{0.627} & \multicolumn{1}{r|}{3879.0} &
\multicolumn{1}{r|}{3271.0} & \multicolumn{1}{r|}{18} &
\multicolumn{1}{r||}{-32} \\
Th (fcc) & \multicolumn{1}{r|}{47.80} & \multicolumn{1}{r|}{48.90} &
\multicolumn{1}{r|}{0.278} & \multicolumn{1}{r|}{1882.8} &
\multicolumn{1}{r|}{2024.0} & \multicolumn{1}{r|}{-7} &
\multicolumn{1}{r||}{43} \\
Tl (bcc) & \multicolumn{1}{r|}{11.00} & \multicolumn{1}{r|}{34.00} &
\multicolumn{1}{r|}{0.309} & \multicolumn{1}{r|}{609.9} &
\multicolumn{1}{r|}{576.0} & \multicolumn{1}{r|}{5} &
\multicolumn{1}{r||}{8} \\
V (bcc) & \multicolumn{1}{r|}{46.00} & \multicolumn{1}{r|}{119.40} &
\multicolumn{1}{r|}{1.228} & \multicolumn{1}{r|}{2444.3} &
\multicolumn{1}{r|}{2178.0} & \multicolumn{1}{r|}{12} &
\multicolumn{1}{r||}{-15} \\
W (bcc) & \multicolumn{1}{r|}{163.10} & \multicolumn{1}{r|}{204.90} &
\multicolumn{1}{r|}{1.005} & \multicolumn{1}{r|}{3497.4} &
\multicolumn{1}{r|}{3653.0} & \multicolumn{1}{r|}{-4} &
\multicolumn{1}{r||}{3} \\  \hline
\end{tabular}
 }
\end{center}
\vskip0.2cm

Table 1.
Theoretical results of melting temperatures (units in K)
derived from the elastic constants of cubic crystals (units in GPa,
gotten from \cite{lb}) compared with the experimental data
(units in K, gotten from
\cite{gschneidner1}). The second-last column shows
the relative error $\kappa=(T _{m,{\rm theor}}/
T _{m,{\rm exp}}-1)\cdot 100$ of our theory, which
is by a factor 2 lower than
the relative error $\delta=(T _{m, {\rm Lind}}/T _{m,{\rm exp}}
-1)\cdot 100$ found from  Lindemann's rule.
The Lindemann melting temperatures $T _{m, {\rm Lind}}$
are from Ref.~\cite{ubl}.
\vskip0.5cm

There is no problem, in principle,
to calculate systematic improvements
to formula (\ref{meltfcc})
by including defect excitations
into the low-temperature approximation (\ref{freelowtfcc})
and stress corrections into the high-temperature
approximation (\ref{htapp}).
This was done in Ref.~\cite{kleinert2}
for the unphysical simple cubic lattices,
and  will be done for bcc and fcc lattices  in future work.
\vskip0.1cm
One of the author (Y.J.) gratefully acknowledges the financial support
from Alexander von Humboldt Foundation.

\end{document}